# Quantum Mechanics as Information Fusion

George Chapline
Lawrence Livermore National Laboratory
Livermore, Ca. 94550

In this letter we provide evidence that quantum mechanics can be interpreted as a rational algorithm for finding the least complex description for the correlations in the outputs of sensors in a large array. In particular, by comparing the self-organization approach to solving the Traveling Salesman Problem with a solution based on taking the classical limit of a Feynman path integral, we are led to a connection between the quantum mechanics of motion in a magnetic field and self-organized information fusion.

**1.** <u>Introduction</u>

Recently it has been pointed out [1] that the intrinsic indeterminism of quantum mechanics is closely analogous to the algorithmic indeterminism of Bayesian networks. In this letter we will argue that this is not accidental, and that actually there is a deep connection between quantum mechanics and the use of Bayesian networks for information fusion [2]. In particular, we will argue that the quantum theory of deviations from a classical path exactly mimics the behavior of a self-organizing Bayesian inference engine.

Actually quantum mechanics seems to correspond most closely to Bayesian inference engines that use a "wake-sleep" algorithm to alternately fix the transition probabilities in a Bayesian network and its dual [3]. This training process minimizes the average information cost of finding the correlations inherent in data sets, and leads to a Boltzmann-like formula relating the probability distribution for various explanations for a given instance of input data to the description costs (in bits) of the various explanations. One class of problems where this approach to information fusion seems particularly promising are situations where the input data represents the outputs from an array of sensors. In these circumstances it is reasonable to assume that the input data is self-organized in the sense that sensors that are in close physical proximity ought to produce similar outputs. It can be shown [4] that in the limit of a large number of sensors the partition function for a self-organized Bayesian network can be recast in the form of a density matrix for the propagation of a quantum mechanical string. This result in itself suggests that the formalism of quantum mechanics is closely related to the self-organization approach to information fusion, but leaves open the question as to whether the complex valued amplitudes of ordinary quantum mechanics have by themselves an information theoretic interpretation.

Our approach to this question is based on an analysis of the Traveling Salesman Problem (TSP). The key point of our analysis is that the TSP can be solved in an obvious way by taking the classical limit of a Feynman path integral. The question whether NP-complete problems like the TSP might be solved using quantum mechanics is, of course, a subject of considerable interest in its own right. What is of interest to us here, though, is how this quantum mechanical approach might be related to the well known approaches to the TSP based on pattern recognition algorithms. For example, good solutions to the TSP can be found using self-organization algorithms [5]. Moreover, the self-organization approach to solving the TSP can be abstractly viewed as an example of information fusion. Thus the TSP problem provides a concrete example of a problem which can be simultaneously viewed as an information fusion problem and a problem in quantum mechanics. It is by comparing these two seemingly very different approaches to the TSP that we are able to perceive a remarkable and heretofore unnoticed connection between quantum amplitudes and self-organized information fusion.

**2.** <u>Helmholtz machines</u>

It has been understood for some time that pattern recognition systems are in essence machines that utilize either preconceived or empirically determined posterior probabilities to classify patterns. In the ideal case where the a priori probability distribution $p(\alpha)$ for the occurrence of various classes $\alpha$ of feature vectors and probability densities $p(x|\alpha)$ for the distribution of data sets $\{x\}$ within each class are known, then the best possible classification procedure would be to simply choose the class $\alpha$ for which the posterior probability

$$P(\alpha \mid x) = \frac{p(\alpha) p(x \mid \alpha)}{\sum_{\beta} p(\beta) p(x \mid \beta)} \tag{1}$$

is largest. Unfortunately in the real world one is typically faced with the situation that neither the class probabilities $p(\alpha)$ nor class densities $p(x|\alpha)$ are precisely known, so that one must rely on empirical information to estimate the conditional probabilities $P(\alpha| x)$ needed to classify data sets.

In practice this has typically meant that one adopts an ad hoc parametric model for the class probabilities and densities, and then uses empirical data to try and fix the parameters of the model. Recently, though, a more elegant approach to this problem, the *Helmholtz machine*, has emerged. The basic idea behind the Helmholtz machine is to use a binary-valued layered Bayesian network together with its dual network as a model for the world. The transition probabilities of the Bayesian network and its dual are fixed using a "wake-sleep" algorithm and the requirement that the length of binary code needed by the "bottom-up" network to explain the various possible states of the world is minimized. The idea that one should always try to find a model for class probabilities and densities that provides the most economical description of the world, due to Rissanen [6], is known as the minimum description length (MDL) principle.

One might think that the best explanation always corresponds to the minimum description cost $E_\alpha$, but this is incorrect because it is possible [3] to devise coding schemes that take advantage of the entropy of alternative explanations for the input data. Instead one wishes to minimize the effective cost $F(x) = \sum_\alpha \{E_\alpha P(\alpha) - (-P(\alpha)\log P(\alpha))\}$ of assigning explanations $\alpha$ to a data set x. In perfect analogy with elementary statistical physics $F(x)$ is minimized when the probabilities of alternative explanations are exponentially related to their description costs by the canonical Boltzmann formula:

$$P(\alpha \mid x) = \frac{e^{-E_\alpha}}{\sum_\alpha e^{-E_\alpha}} \ . \tag{2}$$

Evidently an optimal recognition model should produce a probability distribution $Q(\alpha)$ that is as similar as possible the Boltzmann distribution (2).

What is of particular interest to us in the following is the fact that in the Helmholtz machine both the recognition network and its dual lead to the same probability distribution (2), even though the transition probabilities in the two networks are different. This is very reminiscent of the structure of the Schrodinger picture for quantum mechanics. Indeed, as was first pointed out by Schrodinger himself, quantum mechanics provides a time symmetric description of a continuous Markov process and its dual [7]. Evidently then, one can move a step closer to the set-up of quantum mechanics by replacing the binary valued nodes of the Helmholtz machine by continuous valued nodes.

**3.** Self-organized information fusion

Let us first generalize the notion of a Helmholtz machine so that instead of the state of each node in the network being described by a binary variable, it is a continuous variable $w$, $0 \leq w \leq 2\pi$, that describes some "feature" of the environment. In addition, let us assume that every node on one boundary (corresponding to the "bottom level" of the network) is a measuring device looking at the same environment, and that each node of the network can communicate with a certain number of nodes in the next lower level . As an initial condition for the network we assign to each sensor on the boundary a value $\phi$ that is randomly chosen from a probability distribution for the occurrences of various values of w in the environment, and to each node in the higher layers a value randomly chosen from the interval [0, 2π].

Now intuitively it seems clear that since in principle nearby measuring devices ought to have the similar outputs, a minimal description of the sensor outputs ought to involve just giving the parameters of a smooth curve for $w$ vs. physical location **r** within the network. Therefore it seems reasonable to assume that the Bayesian data processing required for minimal description length information fusion can be modeled by assuming that maps of network nodes into feature space are "self-organizing". If we follow Kohonen's prescription for self-organization [8], this means that the state $w$ of the node located at $\mathbf{r}_i$ will evolve according to a rule of the form

$$w(r,t+1) = w(r,t) + h(r-s)[\phi - w(r,t)], \tag{3}$$

where **s** is the position of the node whose current state $w(\mathbf{s})$ is closest to $\varphi$. the function h(r) is typically assumed to be a Gaussian function peaked at r=0. For simplicity the function h(r-s) can be replaced by the rule that each feature detector is connected to just three of its nearest neighbors. In adopting the rule (3) the data fusion process is modeled as a Markov process whose states are the sets $\{w(\mathbf{r}_i)\}$ of possible states of the feature detectors, and where the transition probabilities are determined by probabilities of occurrence in the environment of various feature orientations $\varphi$. In order to construct an analytical model of this evolution process it will be useful to introduce an energy functional $E[\{w(\mathbf{r}_i)\}]$ such that the self-organization process (3) corresponds to descending the gradient of *E*. Neglecting certain mathematical subtleties, the required energy functional is [9]

$$E[w] = \frac{1}{2} \sum_{<r,s>} \sum_{\phi \in R} P(\phi) |\phi - w(r,t)|^2 \tag{4}$$

where the sum over <r,s> runs over nearest neighbor connections and $R(\mathbf{r})$ is the receptive field of the feature detector located at **r**; i.e. the union of all environmental stimuli closer to $w(\mathbf{r}, t)$ than any other $w(\mathbf{s},t)$, where $\mathbf{s} \neq \mathbf{r}$.

Under the influence of the random variable $\varphi(t)$ the network will relax to an asymptotic state characterized by a stationary probability distribution for various final configurations of feature vectors $\{w(\mathbf{r}_i)\}$. Given the existence of an energy functional, the statistical properties of the stationary set $\{w(\mathbf{r}_i)\}$ can be derived from a "partition function" $Z = \exp[-F(x)]$ which is a sum over all possible stationary state configurations weighted with the Boltzmann factor $\exp(-E[w(\mathbf{r})])$. If we assume that the stochastic evolution of the network is governed by an energy functional of the form (4) then this partition function has the form:

$$Z = \sum_L \kappa^F \prod_{i=l}^{F} \int_0^{2\pi} dw(r_i) e^{-\frac{K}{2} \sum_{<i,j>} |w(r_i) - w(r_j)|^2} \tag{5}$$

where $\kappa$ and *K* are constants, the sum over *L* means a sum over 2-dimensional triangular lattices. When the number of faces N is large, the triangulation *L* can be thought of as approximating a smooth 2-dimensional surfaces, and in the limit N $\to \infty$. It can be shown [10] that the sum over lattices in eq.5 becomes a sum over smooth 2-dimensional surfaces. In this limit the stationary variables *w* ($\mathbf{r}_i$) become a single continuous function of position $\sigma$ on a smooth surface, and the partition function (5) becomes:

$$Z = \int Dw(\sigma) \exp(-S) \tag{6}$$

where the continuum action S is given by

$$S = \frac{K}{2} \int d^2\sigma \partial_\alpha w \partial_\alpha w + S_0 \, . \tag{7}$$

The constant $S_0$ in (7) replaces the constant $\kappa$ in eq. 5. If we generalize the angle variable $w$ to a variable taking values in the complex plane, then the partition function (6) would have an interesting physical interpretation [10]; namely, it would represent the density matrix for a "string" moving on a smooth 2-dimensional surface. In this string interpretation one can think of the magnitude of $w(\sigma)$ as representing the local magnification of the mapping from the network into feature space.

It is now clear why self-organization provides a natural solution to the problem of finding MDL explanations for the outputs of a large number of measuring devices. In the large N limit the "explanations" will be represented by smooth mappings into feature space of a finite 2-dimensional surface representing the physical layout and connectivity of an information fusion network associated with a large number of measuring devices. The information cost of any particular explanation will be just the (finite) quantized action given in eq. 7. The natural unit of quantization, i.e. the surface area in feature space equivalent to 1 bit, is determined by the inverse of the constant $K$ in eq. 7. Taking into account the information cost savings of entropic coding, the explanation cost averaged over all possible explanations will for large N be just the negative logarithm of the string partition function Z defined in eq. 7.

Although the string action (7) obviously has a natural tendency to minimize the area in feature space that represents the information cost of an explanation, it remains to be clarified how the density matrix formally defined in eq. 7 can account for the compression of raw sensor data into simple characterizations of the environment. Indeed what one normally wants to do with sensor data is to confirm or falsify some simply hypothesis concerning the state of the environment. In the following we focus on the TSP as a simple model for self-organized information fusion involving data compression.

**4.** The Traveling Salesman Problem and quantum mechanics

The self-organization approach to solving the TSP, originally due to Durbin and Willshaw [11], makes use of an "elastic ring" consisting of movable nodes with elastic bands connecting each node. Solving the TSP essentially involves stretching the elastic bands in such a way that the ring intersects each city once and only once, in which case the ring represents a global tour through the cities. In practice, one tries to minimize the cost function:

$$E(\{w_i\}) = -\lambda^2 \sum_\mu \log \sum_i \left[ \exp\left(\frac{|\xi^\mu - w_i|^2}{2\lambda^2}\right) + \frac{K}{2}|w_{i+1} - w_i|^2 \right] \tag{8}$$

where $w_i$ represents the position of the i-th node, $\xi^\mu$ represents the position of the μ-th city., $\lambda$ is the range of attraction for the cities, and the $K$ term is just the energy functional for self-organization introduced in eq.5 of the last section. Typically the range $\lambda$ is assumed to decrease with time. In the limit where the $w_i$ approach the $\xi^\mu$ and $\lambda \to 0$, the first term approaches zero and the second term is minimized by the shortest possible path between cities.

Durbin and Willshaw showed that If the parameters in eq. (8) are properly chosen, then descending the gradient of $E(\{w_i\})$ will generally provide an acceptable solution to the TSP. Indeed the exponential terms on the r.h.s. of eq. 8 force every point $w_l(t)$ to approach towards some city $\zeta^\mu$, while the second term acts to minimize the total length of the elastic ring. As a result of their focus on the difficulty of finding solutions to the TSP when the number of cities is large, Durbin and Willshaw had in mind spreading out the initial points $w_i(0)$ so they would be attracted to a large number of cities, thereby solving an effectively NP complete problem. For our purposes, though, we are quite content to consider situations where the initial points may be attracted to only a few cities as illustrated in Fig. 1.

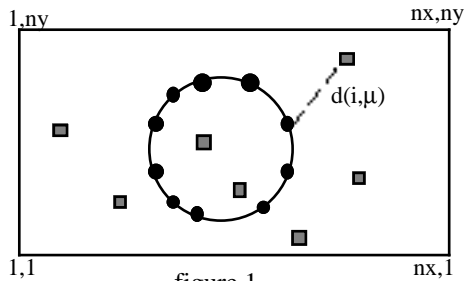

figure 1.

As indicated in the figure, the initial point $w_l(0)$ will be attracted to the city $\zeta^\mu$ if the $d(i,\mu)$ is less than or on the order of the range $\lambda(0)$. As the points $w_l(t)$ approach their final destinations the ring of elastic bands connecting these point will sweep out a surface. If one thinks of the set of initial points $\{w_i(0)\}$ as the input feature data for a Helmholtz machine, then in accordance with our previous discussion one might adopt the area of the surface swept out by the elastic ring as a measure of the information cost of arriving at the "explanation" for the input feature data.

It is self-evident that the TSP can also be solved, at least in principle, by the expedient of constructing a Feynman path integral such that a) all the paths are closed and constrained so that each path passes at least once through a certain set of "cities" $\zeta^\mu$, b) the action is taken to be the path length, and c) the classical limit is taken by letting `h` -> 0. It is not our purpose here to pursue such a program for solving the TSP, but instead to compare the quantum path integral approach to solving the TSP with the Durbin-Willshaw approach. At first sight the two approaches don't seem to have much in common. However, one of the remarkable insights into the nature of quantum mechanics provided by Feynman's derivation of his path integral is that the fractal dimension of typical quantum paths for a non-relativistic particle is 2 [12]. As a consequence, if we restrict the quantum paths being considered for the TSP to lie inside the area swept out by the Durbin-Willshaw elastic ring, then typical paths will pretty much fill up this area. Thus we may anticipate that there may be some connection between the 2-dimensional action (7) and a path integral over constrained quantum paths for a non-relativistic quantum particle.

In fact there is an interesting connection between the quantum mechanical motion of a non-relativistic particle constrained to pass through certain points and the string action, eq. (7). To see this we first note that the

Feynman path integral for free particle propagation between two cities $\zeta^a$ and $\zeta^b$ in a time $t_b - t_a$ can be written in the form

$$K(\zeta^b, t_b; \zeta^a, t_a) = F(t_b - t_a) \exp(IS_{cl}), \qquad (9)$$

where $S_{cl} = m(\zeta^b - \zeta^a)^2 / 2(t_b - t_a)$ is the classical action and $F(t_b - t_a)$ is a path integral over deviations from the classical path. This path integral has the form:

$$F(t_b - t_a) = \int \exp\left[\frac{i}{\hbar}\int_{t_a}^{t_b} \frac{m}{2}(\dot{x} - v(t))^2 dt\right] D\mathbf{y}(t'), \qquad (10)$$

where $\mathbf{y} = \mathbf{x}(t) - \mathbf{x}_c(t)$ is the deviation from the classical path $v(t)$ is a specified velocity along the classical path, and $y(t_b) = y(t_a) = 0$. Now it is a amusing fact that the action (10) can be transformed into the same form as the action for a non-relativistic charged particle in a magnetic field. In particular, since $v(t)$ must be defined for all positions $\mathbf{x}$, we can introduce a vector potential $v(t) = (1/m) \mathbf{A}(\mathbf{x},t)$. If we now make a gauge transformation and choose a gauge function that satisfies the Hamilton-Jacobi equation, it can be shown [13] that the action function in (10) can be written in the form:

$$S(\mathbf{x}) = \frac{1}{\hbar}\int_{t}^{t_b}\left[\frac{m}{2}\dot{x}^2 + \dot{x}A(x)\right]dt. \qquad (11)$$

In other words the free particle action associated with deviation from a prescribed classical path can be rewritten as the action for a particle moving in a magnetic field. If the deviations from the classical path between $\zeta^a$ and $\zeta^b$ are restricted to lie in a region whose length is $|\zeta^b - \zeta^a|$ and average width is $d_y$, then by well known arguments [14] the number N of quantum states corresponding to the first Landau level is $A H / 2\pi \hbar$, where $A = |\zeta^b - \zeta^a| d_y$ is the area of the region in which deviations from the classical path are allowed and $\mathbf{H} = \nabla \times \mathbf{A}(\mathbf{x},t)$ is the effective magnetic field. If we assume that the first Landau level is filled and the particles are non-interacting, the wavefunction for the first Landau level will have the form:

$$\Psi = \prod_{i=1}^{N} e^{-\frac{(z-z)^2_i}{4}}, \qquad (12)$$

where the set $\{z_i\}$ is uniformly distributed over the area $A$. We see therefore that the wavefunction for the first Landau level can be written in the form $\exp(-A/A_0)$ where $A_0$ is an elementary area on the order of $H/2\pi\hbar$.

Thus we arrive at our basic result that the quantum mechanical wavefunction describing deviations from a classical path through fixed points $\xi^\mu$ has the same form as the MDL conditional probability for arriving at these same points starting from almost arbitrary initial values. Because of the important role played by the area swept out by the Durbin-Willshaw elastic ring, one might say, as has been previously suggested [15], that quantum mechanics has its origin in a fundamental relationship between information and area. A perhaps more interesting way to look at this relationship, though, is to say that quantum mechanics is really all about providing a MDL representation for the correlations inherent in the outputs from a large number of measuring devices.

5. <u>Some remarks</u>

It is of course tempting to identify the quantized area in (12) with an information cost. Because this is an area in physical feature space rather than phase space one might think that this quantization condition is different than the familiar Bohr-Sommerfield phase space quantization. As it happens though, because for a particle moving in a magnetic field $d_y$ is related to $\Delta p_x$, our area quantization condition is just the usual phase space quantization condition in disguise. Thus our identification of the area in (12) with an information cost is consistent with old ideas for relating phase space quantization to information. This idea was first put forward in the 1920's, when it was recognized that a rule of thumb to divide phase space into "cells" each of whose volume was equal to $h^f$, where $f$ is the number of degrees of freedom could solve the long-standing problem of calculating the entropy constant of a gas [16].

Since there is nothing intrinsically mysterious about Bayesian probabilistic reasoning, one might hope to exploit the connection we have found between Feynman's path integral formulation of quantum mechanics and Bayesian information fusion to better understand the non-intuitive aspects of quantum mechanics. The general relationship between the indeterminism of quantum mechanics and Bayesian decision trees was already commented on in ref. 1. In the case of Helmholtz machines though it appears that one can go much further in finding Bayesian interpretations for the peculiar features of quantum mechanics. For example, spatially non-local correlations between the results of measurements carried out at physically separate locations naturally occur at the input level of a Helmholtz machine. One way to understand this is to note that data input at a node of the lowest input level changes the probability distribution for the state of all higher level network nodes that are connected to the lowest level node via bottom-up connections. The influence on higher level nodes then propagates backward via the top-down transition probabilities to affect the probability distribution at other input level nodes.

Whether the spatially non-local correlations present in the bottom level of a Helmholtz machine are precisely of the right form to explain Bell's inequalities and other mysterious features of quantum entangled states remains to be seen. However, it seems quite possible that these mysteries will eventually yield to our view of the fundamental nature of quantum mechanics as a rational basis for combining together the outputs of multiple sensors.

Acknowledgment: The author is very grateful for discussions with Jim Barbieri.